\documentclass[runningheads]{llncs}
\pdfoutput=1
\usepackage[utf8]{inputenc}
\usepackage{graphicx}
\usepackage{amsmath}
\usepackage{amssymb}
\usepackage{multirow}
\usepackage{makecell}
\usepackage{tabularx} 
\usepackage{dblfloatfix}
\usepackage[shortlabels]{enumitem}
\setlist[enumerate]{nosep}

\usepackage{array, caption, floatrow, tabularx, makecell, booktabs}%
\setcellgapes{3pt}
\usepackage{booktabs}
\usepackage{array}
\usepackage{adjustbox}

\usepackage{floatrow}
\newcommand{\mytoprule}{\specialrule{0.1em}{0em}{0.33em}}
\newcommand{\mybottomrule}{\specialrule{0.1em}{0.1em}{0em}}

\makeatletter
\def\@fnsymbol#1{\ensuremath{\ifcase#1\or \dagger\or \star \else\@ctrerr\fi}}
\makeatother
   
\begin{document}

\title{Automatic ultrasound vessel segmentation with deep spatiotemporal context learning}

%

\author{
Baichuan Jiang\inst{1,2,}\thanks{These authors contributed equally.} \and
Alvin Chen\inst{1,\dagger} \and 
Shyam Bharat\inst{1} \and
Mingxin Zheng\inst{1}
}

\authorrunning{B. Jiang et al.}

\titlerunning{Automatic vessel segmentation with deep spatiotemporal context learning}
\institute{Philips Research North America, 222 Jacobs St. Cambridge, MA 02141, USA \email{\{alvin.chen, shyam.bharat, mingxin.zheng\}@philips.com} \and \thanks{Work done during internship at Philips Research.}Department of Computer Science, Johns Hopkins University, Baltimore, MD 21218, USA \email{bcjiang@jhu.edu} }
\maketitle              
\begin{abstract}
Accurate, real-time segmentation of vessel structures in ultrasound image sequences can aid in the measurement of lumen diameters and assessment of vascular diseases. This, however, remains a challenging task, particularly for extremely small vessels that are difficult to visualize. We propose to leverage the rich spatiotemporal context available in ultrasound to improve segmentation of small-scale lower-extremity arterial vasculature. We describe efficient deep learning methods that incorporate temporal, spatial, and feature-aware contextual embeddings at multiple resolution scales while jointly utilizing information from B-mode and Color Doppler signals. Evaluating on femoral and tibial artery scans performed on healthy subjects by an expert ultrasonographer, and comparing to consensus expert ground-truth annotations of inner lumen boundaries, we demonstrate real-time segmentation using the context-aware models and show that they significantly outperform comparable baseline approaches.

\end{abstract}
\section{Introduction}
Over 120 million people worldwide are affected by peripheral vascular disease, making it one of the leading causes of morbidity and mortality globally \cite{pvd_ref}. 
Vascular ultrasound (US) based on B-mode and Color Doppler imaging is widely used to evaluate luminal narrowing and flow in stenotic vessel segments. Accurate delineation of vessel wall boundaries is crucial for diagnosis and disease prognosis \cite{moccia_2018_review}, but can be highly challenging due to complex vascular anatomy, visual ambiguity, acoustic imaging artefacts, probe motion, and very small target structures hidden within the US image frame \cite{liu_2019_review}. 
Automated, real-time segmentation of relevant vessels can aid in clinical assessment while providing a means to improve vascular US imaging workflows and reduce operator dependency. 

Prior work on US vessel segmentation have utilized shape and motion models \cite{guerrero_2007,ma_2018,patwardhan_2012,mistelbauer_2021} typically requiring initialization with seed points in the first frame. To provide additional context, the inclusion of flow information alongside the B-mode image has been proposed \cite{keil_2012,tamimi_2017,moshavegh_2016,akkus_2015}. In recent years, deep learning has received increased focus \cite{moccia_2018_review,smistad_2016}, with the majority of work on vessel segmentation utilizing UNet/VNet-like models operating on individual frames \cite{zhou_2021,zhou_2019}. Incorporating the time dimension, recurrent mechanisms have been combined with convolutional networks as a way to encode temporal memory from image sequences \cite{gao_2019,milletari_2018,arbelle_2018,webb_2020,duque_2020,mirunalini_2019}. However, to date, most studies have focused on segmentation of large carotid \cite{zhou_2021,zhou_2019} or coronary \cite{mirunalini_2019} arteries using UNet/VNet, and accuracy on small extremity vasculature has not been systematically investigated.

\begin{figure}
\includegraphics[width=0.82\textwidth]{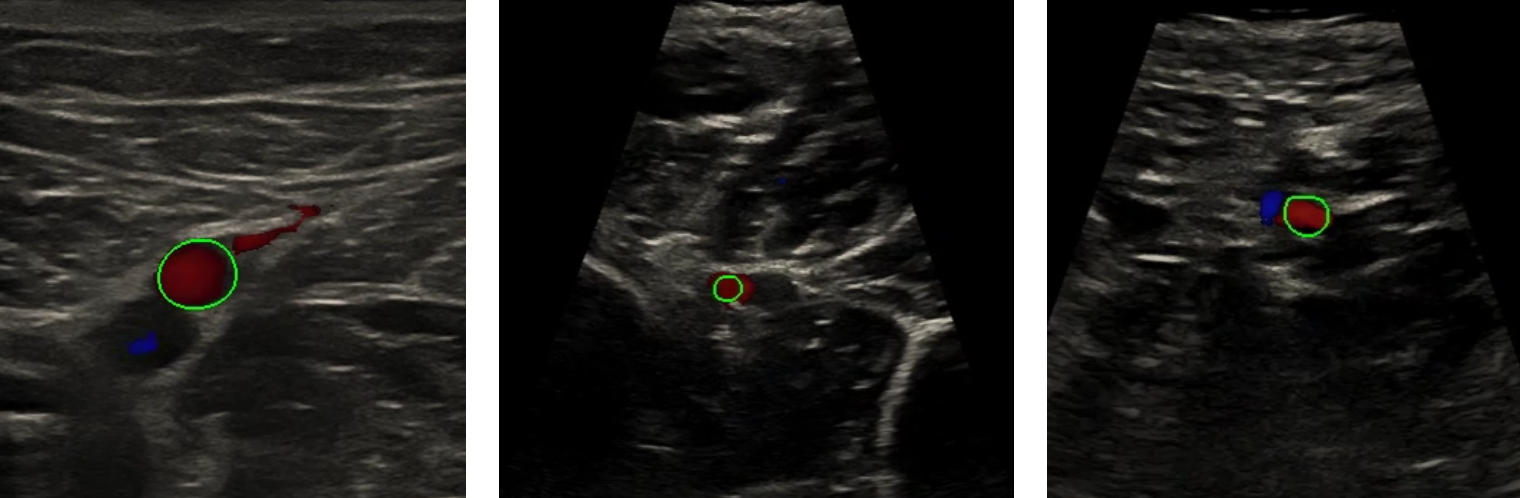}
\caption{B-mode and Color Doppler US images of lower-extremity arterial vasculature. Expert ground-truth annotations of inner lumen vessel wall boundaries shown in green. \textit{Left}: femoral artery ($\sim$5 mm diameter), \textit{Middle}: anterior tibial artery ($\sim$2 mm diameter), \textit{Right}: posterior tibial artery ($\sim$3 mm diameter).}\label{fig_vessel_type}
\end{figure}

\vspace{-16pt}
\subsubsection{Our Contributions.}
We demonstrate that a spatiotemporally-aware deep learning model is capable of automatic, real-time segmentation of inner lumen vessel boundaries from challenging freehand US sequences  (Fig. \ref{fig_vessel_type}). Improved performance in difficult anatomy, namely small-scale lower-extremity peripheral arterial vasculature, is made possible by leveraging the rich spatiotemporal information available in US and utilizing dual-input B-mode and color flow signals. Our approach aims to simulate the contextual inferencing processes of experienced ultrasonographers, who are trained to recognize temporal signatures in both modalities and attend to small structures of interest while ignoring background. 

Specifically, we propose a fully convolutional encoder-decoder network that feeds B-mode and Color Doppler inputs through a series of spatial, temporal, and channel-wise contextual units embedded within each resolution layer. 
By preserving multi-resolution features as they pass through each unit, the network is able to propagate learned representations temporally across all scales. 

\textit{Additional contributions of the paper}:
\begin{enumerate}
  \item We systematically study the impact of the multi-scale spatial, temporal, and channel-wise embeddings for segmentation of femoral (4-6 mm diameter) and tibial (2-3 mm diameter) arteries in a series of network ablation experiments.
  \item We investigate the utility of exploiting combined B-mode and Color Doppler information compared to B-mode alone, and we evaluate the benefits of domain-specific augmentation on the US small-structure segmentation task.
  \item We compare segmentation results against consensus ground-truth annotations from multiple clinical experts, and we demonstrate significant improvements in accuracy using the proposed methods compared to baseline models. 
\end{enumerate}

\section{Methodology}
\subsubsection{Model overview.} 
\begin{figure}
\includegraphics[width=\textwidth]{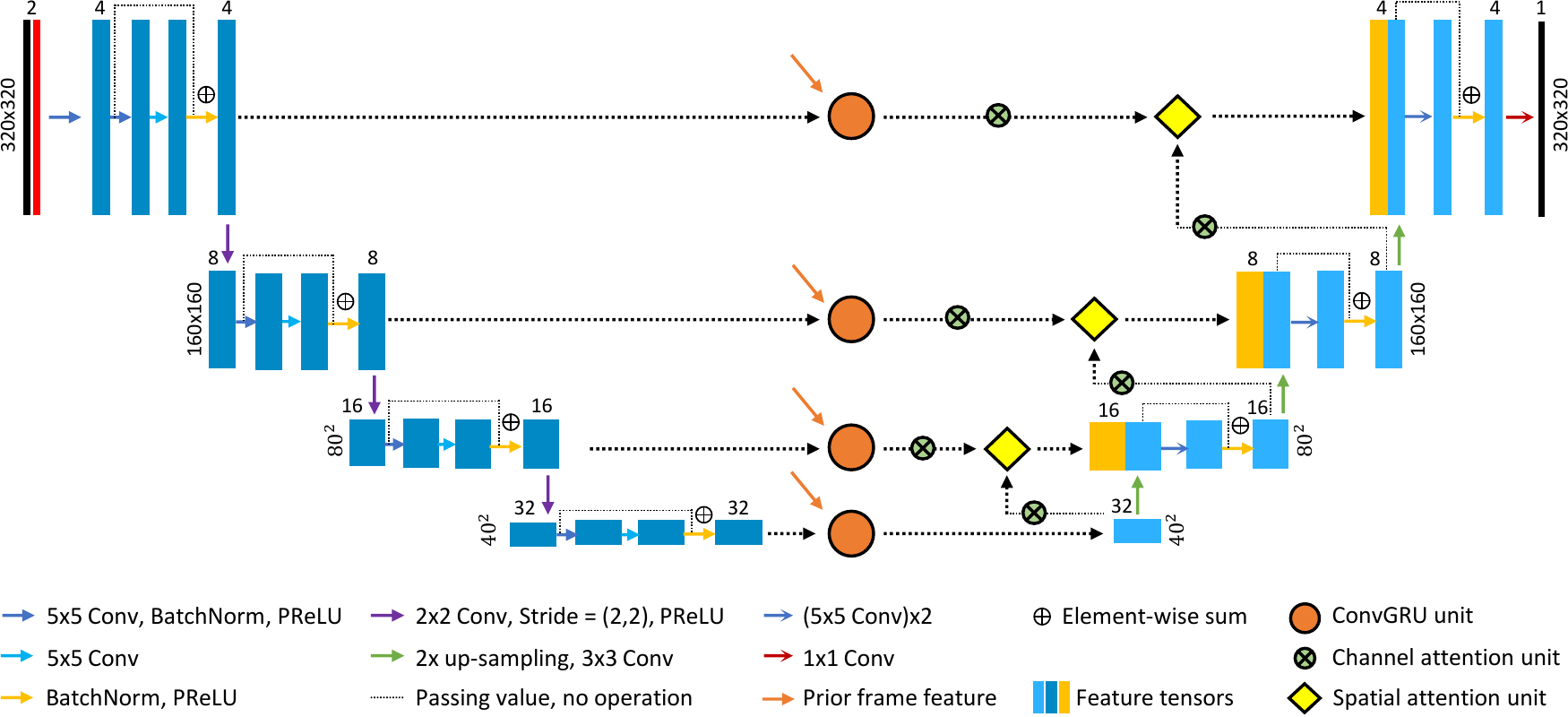}
\caption{Network architecture. The model feeds B-mode and Color Doppler inputs through a multi-scale series of spatial, temporal, and feature-wise contextual units embedded in each resolution layer of a fully convolutional encoder-decoder backbone. The design enables efficient learning of multi-modal spatiotemporal context information for challenging small-structure segmentation tasks.} \label{fig_net}
\end{figure}

The proposed VESsel NETwork with Spatial, Channel and Temporal context (VesNetSCT+) is illustrated in Fig. \ref{fig_net}. The network architecture and implementation details are described below.

\vspace{-16pt}
\subsubsection{Encoder-Decoder Backbone.}
Our model uses a UNet/VNet-like\cite{unet_ref,vnet_ref} backbone with a two-channel input to support B-mode and Color Doppler frames. The proposed spatial, temporal, and feature/channel-wise contextual units are sequentially embedded along the residual skip connections\cite{resnet_ref} in each resolution layer. This design allows the model to aggregate learned representations across multiple spatial scales and time points. We additionally posit that, unlike methods which explicitly modify the backbone network \cite{gao_2019,milletari_2018,arbelle_2018,webb_2020}, embedding into the skip connections minimizes disruption of gradients along the main path. 

\textit{Implementation details:} B-mode and Color channels are normalized between 
(0,1) and (-1,1), respectively. We apply batch normalization and PReLU, and we use resize-up-convolutions \cite{resize_conv_ref} in the decoder to minimize checkerboard artifacts. We also reduce the total model size from 30M (original UNet) to 0.3M by opting for a four-layer network backbone capable of real-time inference on clinical US machines. Networks are trained via RMSProp with initial learning rate of 0.0001.



\subsubsection{Multi-scale Temporal Gating.}  
To introduce temporal context, a hierarchical series of convolutional gated recurrent units (ConvGRU) \cite{convgru_ref,convgru0_ref} are embedded between the contracting and expanding paths (round orange blocks in Fig. \ref{fig_net}). By replacing dot product operations found in standard GRU with convolutional operations, the learned temporal representations have the inherent spatial connectivity of convolutional networks and are more parameter-efficient \cite{convgru0_ref}. 

\textit{Implementation details:} The inputs into the ConvGRU are the feature maps $x_t$ for the current frame $t$ and the hidden states $h_{t-1}$ from the prior time point: 
\begin{eqnarray}
z_t &=& \sigma(W_{hz}\ast h_{t-1} + W_{xz}\ast x_{t} +b_z) \\
r_t &=& \sigma(W_{hr}\ast h_{t-1} + W_{xr}\ast x_{t} +b_r) \\
\hat{h_t} &=& \tanh(W_h\ast (r_t \odot h_{t-1}) + W_x\ast x_t + b) \\
h_t &=& (1-z_t)\odot h_{t-1} + z\odot \hat{h_t}
\end{eqnarray}
where $W_{hz}, W_{xz}, W_{hr}, W_{xr}, b_z, b_r, b$ are learnable convolution kernels and biases. $z_t$ and $r_t$ are internal update and reset gates. The output $h_{t}$ is computed as a weighted sum of $h_{t-1}$ and candidate activation $\hat{h_t}$. 
$\sigma$ denotes sigmoid activation, $\odot$ denotes element-wise multiplication, and $\ast$ denotes convolution. All temporal models are trained via truncated backpropagation-through-time (TBTT) with a fixed sequence length of 50 frames. A time window of 1-4 frames is used in the feedforward step and accumulated to compute back-propagation gradients \cite{tbtt_ref}.

\subsubsection{Multi-scale Spatial Attention.}  
We add soft self-attention gates to provide higher-resolution layers with spatial support from coarse layers.
This idea has been shown to improve accuracy on small-structure segmentation tasks \cite{attention_unet_ref} and is fully differentiable, unlike hard attention mechanisms based on iterative region proposal and cropping \cite{mnih_2014}. The spatial self-attention units are introduced in a multi-scale manner along the skip connections, immediately following the temporal and channel operators. The arrangement provides the self-attention gates with access to global spatiotemporal context when learning to suppress irrelevant background and attend to regions of interest.

\textit{Implementation details:} Within this unit, input feature maps $x$, and input gating feature maps $g$ from the previous coarse resolution level, are used to derive attention multiplier maps $\alpha \colon\left[ 0,1 \right] \to \mathbb{R}^2$. The self-attention maps are multiplied element-wise into each channel of $x$ to produce the output maps $\hat{x}$:
\begin{eqnarray}
\alpha &=& \sigma(\psi(\delta(W_x \ast x + W_g \ast g +b_g)) + b_\psi) \\
\hat{x_c} &=& x_c \odot \alpha
\end{eqnarray}
where $\sigma$ and $\delta$ denote sigmoid and ReLU activations, respectively. $W_x,W_g,\psi$ are weight parameters for the channel-wise $1\times1\times1$ convolutions, $b_\psi,b_g$ are the bias terms in the gating unit, and $\odot$ denotes element-wise multiplication between the multiplier and input feature maps at channel $c$. 

\subsubsection{Multi-scale Feature-wise Channel Attention.}
Finally, feature/channel self-attention is incorporated in the skip connections and along the decoder path. Specifically, we employ convolutional block attention modules \cite{cbam_ref} immediately before each spatial attention gate (round green blocks in Fig. \ref{fig_net}). Particularly for tasks involving multiple input modalities, self-attention along the channel dimension offers a mechanism to explicitely model interdependencies between modalities (in our case, the encoded B-mode and Color Doppler input signals).

\textit{Implementation details:} Given input feature maps $x^c$ with $c$ channels, average- and max-pooling are performed along the channel dimension to produce descriptors $x^c_{avg},\  x^c_{max}\in \mathbb{R}^{1\times 1\times C}$. These are passed through a shared multi-layer perceptron (MLP) $g$ and element-wise summed. The result, a vector $m_c\in \mathbb{R}^{1\times 1\times C}$, is channel-wise multiplied ($\otimes$) with the input $x^c$: 
\begin{eqnarray}
\hat{x}^c &=& m_c \otimes x^c  \\
      &=& \{\sigma (g(x^c_{avg}) + g(x^c_{max}))\} \otimes x^c   \\
      &=& \{\sigma (W_1(\delta(W_0(x^c_{avg}))) + W_1(\delta(W_0(x^c_{max}))))\} \otimes x^c
\end{eqnarray}
where $\sigma$ and $\delta$ denote sigmoid and ReLU, and $W_0, W_1$ are learnable MLP weights.

\subsubsection{Domain-specific Augmentation.}
Extensive data augmentation was applied in training to improve generalizability and robustness to real-world freehand US imaging conditions. Augmentations were defined on training sequences spanning 50 sequential frames, and included: (1) \textit{Spatial augmentation} based on random translation, rotation, scaling, cropping, and horizontal flipping; (2) \textit{Gain/contrast augmentation} based on random adjustment of histogram and time gain compensation curves separately to the B-mode and Color channels; (3) \textit{Color Doppler augmentation}, where channel dropout is applied with fixed probability on the Color inputs to simulate poor Doppler signal due to impaired blood flow; and (4) \textit{Temporal augmentation} by varying the start and end frames, interval between frames, and order of frames (forward or reverse) for each set of 50-frame inputs.

\subsubsection{Vascular Ultrasound Data Acquisition.}
Freehand lower-extremity arterial US exams were performed by an expert vascular sonographer on left and right legs of 7 healthy subjects. Scans were acquired in transverse orientation following standard imaging workflows for diagnostic Duplex ultrasonography \cite{duplex_ref}. The scans were performed under simultaneous B-mode and Color Doppler modes and spanned the length of the leg, from ankle to groin. In total, 22 exam sequences were acquired, including 13 along the femoral arteries and 9 along the anterior/posterior tibial arteries. All data were collected with a Philips Epiq 7 system and 12 MHz linear transducer (pixel spacing $\sim$0.1mm, frame rate $\sim$15Hz).

\subsubsection{Clinical Expert Ground-Truth Annotation.}
Expert ground-truth annotations of inner lumen boundaries (Fig. \ref{fig_vessel_type}) were provided by two experienced vascular sonographers. Consensus ground-truth masks were computed via shape-based averaging of the two sets of individual segmentations \cite{rohlfing_2005}. A total of 30,839 consensus-annotated frames were obtained from the 22 femoral and tibial artery exams in this manner. The data were divided according to subject to allow for independent training and testing, and to perform leave-one-out cross-validation.



\section{Results}

\begin{table}[!b]
\floatsetup{floatrowsep=qquad, captionskip=4pt}
\addtolength{\parskip}{-1em}
\setlength{\tabcolsep}{0.7pt}
{\caption{Summary of segmentation performance on femoral and tibial test sets. 
}
  \label{tab_summary}}
\begin{center}
\begin{tabular*}{\textwidth}{ccccccc} 
 \mytoprule
  \makecell{\textbf{Model}\\\textbf{name*}} & 
  \makecell{\textbf{\#}\\\textbf{params}} & 
  \makecell{\textbf{Input}\\\textbf{channels}}   & 
  \makecell{\textbf{Spatial/}\\\textbf{channel}\\\textbf{attention}} & 
  \makecell{\textbf{Temporal}\\\textbf{gating}} & 
  \makecell{\textbf{Time}\\\textbf{window}} & 
  \makecell{\textbf{Dice score}\\\textbf{(mean$\pm$std)}} \\
 \midrule
 \midrule
 \multicolumn{7}{c}{Femoral arteries (4-6 mm diameter)} \\
 \midrule
 Baseline     & 103k  & Bmode        & -     & -            & -  & 0.775 $\pm$ 0.282\\
 Baseline-L   & 310k  & Bmode        & -     & -            & -  & 0.788 $\pm$ 0.301\\
 \cmidrule(lr){2-2}\cmidrule(lr){3-3}\cmidrule(lr){4-4}\cmidrule(lr){5-5}\cmidrule(lr){6-6}\cmidrule(lr){7-7}
 VesNet       & 103k  & Bmode+Color  & -     & -            & -  & 0.870 $\pm$ 0.150\\
 VesNetS      & 105k  & Bmode+Color  & S     & -            & -  & 0.881 $\pm$ 0.175\\
 VesNetSC     & 106k  & Bmode+Color  & S+C   & -            & -  & 0.903 $\pm$ 0.101\\
  VesNet-L      & 313k  & Bmode+Color  & -     & -            & -  & 0.887 $\pm$ 0.180\\
 \cmidrule(lr){2-2}\cmidrule(lr){3-3}\cmidrule(lr){4-4}\cmidrule(lr){5-5}\cmidrule(lr){6-6}\cmidrule(lr){7-7}
 VesNetT      & 259k  & Bmode+Color  & -     & Single       & 1  & 0.908 $\pm$ 0.086\\
 VesNetT+     & 307k  & Bmode+Color  & -     & Multi-scale  & 1  & 0.914 $\pm$ 0.065\\
 VesNetST+    & 309k  & Bmode+Color  & S     & Multi-scale  & 1  & 0.919 $\pm$ 0.069\\
 VesNetSCT+   & 310k  & Bmode+Color  & S+C   & Multi-scale  & 1  & 0.925 $\pm$ 0.051\\
 VesNetSCT++  & 310k  & Bmode+Color  & S+C   & Multi-scale  & 4  & \textbf{0.927 $\pm$ 0.041}\\
 \midrule
 \midrule
 \multicolumn{7}{c}{Tibial arteries (2-3 mm diameter)} \\
 \midrule
 Baseline     & 103k  & B-mode        & -     & -            & -  & 0.133 $\pm$ 0.227\\
 Baseline-L   & 310k  & B-mode        & -     & -            & -  & 0.164 $\pm$ 0.254\\
 \cmidrule(lr){2-2}\cmidrule(lr){3-3}\cmidrule(lr){4-4}\cmidrule(lr){5-5}\cmidrule(lr){6-6}\cmidrule(lr){7-7}
 VesNet       & 103k  & Bmode+Color  & -     & -            & -  & 0.527 $\pm$ 0.336\\
 VesNetS      & 105k  & Bmode+Color  & S     & -            & -  & 0.564 $\pm$ 0.317\\
 VesNetSC     & 106k  & Bmode+Color  & S+C   & -            & -  & 0.570 $\pm$ 0.282\\
 VesNet-L      & 313k  & Bmode+Color  & -     & -            & -  & 0.534 $\pm$ 0.328\\
 \cmidrule(lr){2-2}\cmidrule(lr){3-3}\cmidrule(lr){4-4}\cmidrule(lr){5-5}\cmidrule(lr){6-6}\cmidrule(lr){7-7}
 VesNetT      & 259k  & Bmode+Color  & -     & Single       & 1  & 0.564 $\pm$ 0.283\\
 VesNetT+     & 307k  & Bmode+Color  & -     & Multi-scale  & 1  & 0.655 $\pm$ 0.211\\
 VesNetST+    & 309k  & Bmode+Color  & S     & Multi-scale  & 1  & 0.664 $\pm$ 0.246\\
 VesNetSCT+   & 310k  & Bmode+Color  & S+C   & Multi-scale  & 1  & 0.671 $\pm$ 0.240\\
 VesNetSCT++  & 310k  & Bmode+Color  & S+C   & Multi-scale  & 4  & \textbf{0.679 $\pm$ 0.195}\\
 \mybottomrule
\end{tabular*}
\end{center}
* \textit{Nomenclature}. Baseline: UNet; VesNet: bimodal-input UNet; ``-L": larger network with more channels per layer; ``S": spatial attention; ``C": channel attention; ``T":  temporal gating; ``+": multi-scale embeddings; ``++": expanded TBTT window.
\end{table}
\addtolength{\parskip}{-0.05em}

\subsubsection{Validation $\&$ Network Ablation Studies.}\hfill\\
\textit{Temporal, spatial, and channel-wise context:} On both femoral and tibial hold-out test data, we saw significant improvements in Dice scores (Table \ref{tab_summary}) with the addition of temporal, spatial, and feature/channel-wise contextual embeddings alongside bimodal B-mode+Color inputs. For the femoral dataset, the best-performing context-aware model (VesNetSCT++, $0.927\pm 0.041$ Dice) demonstrated an improvement of 18 to 20\% in comparison to two baseline UNet models of varying sizes (Baseline, $0.775\pm 0.282$ Dice; Baseline-L, $0.788\pm 0.301$ Dice). Meanwhile, in the tibial arteries, which represented $<$0.1\% of pixels in each frame, the baseline models failed entirely (Baseline, $0.133\pm 0.227$ Dice; Baseline-L, $0.164\pm 0.254$ Dice) compared to the context-aware model (VesNetSCT++, $0.679\pm 0.195$ Dice).


\textit{Multi-scale embeddings:} The introduction of temporal/spatial/channel embeddings in a multi-scale manner (VesNetT+) resulted in improved performance compared to equivalent models with the embeddings applied only to the innermost layer of the encoder-decoder backbone (VesNetT), as proposed in \cite{gao_2019,milletari_2018,arbelle_2018,webb_2020}.

\textit{Temporal window:} We experimented with the time window for accumulating feed-forward and back-propagation updates when training temporal models using TBTT \cite{tbtt_ref}. Improvements were seen by increasing the window size (VesNetSCT++, time window=4), at the cost of longer training times.

\begin{table}[!b]
  \floatsetup{floatrowsep=qquad, captionskip=4pt}
  \begin{floatrow}[2]
    \ttabbox%
    {\begin{tabularx}{0.465\textwidth}{cccc}
      \mytoprule
      \makecell{\textbf{Model}\\\textbf{name}} & 
      \makecell{\textbf{Color}\\\textbf{Doppler}\\\textbf{dropout}} & 
      \makecell{\textbf{Dice}\\\textbf{score}}\\
      \midrule
      \midrule
      \multicolumn{3}{c}{Femoral arteries (4-6 mm diameter)} \\
      \midrule
      VesNetSC   & 0.0 & 0.880 $\pm$ 0.145\\
      VesNetSC   & 0.4 & 0.903 $\pm$ 0.101\\
      \cmidrule(lr){2-2}\cmidrule(lr){3-3}
      VesNetSCT+ & 0.0 & 0.911 $\pm$ 0.087\\
      VesNetSCT+ & 0.4 & \textbf{0.925 $\pm$ 0.051}\\
      \midrule
      \midrule
      \multicolumn{3}{c}{Tibial arteries (2-3 mm diameter)} \\
      \midrule
      VesNetSC   & 0.0 & 0.579 $\pm$ 0.312\\
      VesNetSC   & 0.4 & 0.570 $\pm$0.282\\
      \cmidrule(lr){2-2}\cmidrule(lr){3-3}
      VesNetSCT+ & 0.0 & \textbf{0.676 $\pm$ 0.293}\\
      VesNetSCT+ & 0.4 & 0.671 $\pm$ 0.240\\
      \mybottomrule
      \end{tabularx}}
    {\caption{Impact of color augmentation. See Table 1 nomenclature.}
      \label{tab_color_augm}}
    \hfill%
    \ttabbox%
    {\begin{tabularx}{0.465\textwidth}{cccc}
      \mytoprule
      \makecell{\textbf{Model}\\\textbf{name}} & 
      \makecell{\textbf{Temporal}\\\textbf{augm-}\\\textbf{entation}} & 
      \makecell{\textbf{Dice}\\\textbf{score}}\\
      \midrule
      \midrule
      \multicolumn{3}{c}{Femoral arteries (4-6 mm)} \\
      \midrule
      VesNetSCT+ & No  & 0.917 $\pm$ 0.117\\
      VesNetSCT+ & Yes & \textbf{0.925 $\pm$ 0.051}\\
      \midrule
      \midrule
      \multicolumn{3}{c}{Tibial arteries (2-3 mm)} \\
      \midrule
      VesNetSCT+ & No  & 0.660 $\pm$ 0.245\\
      VesNetSCT+ & Yes & \textbf{0.671 $\pm$ 0.240}\\
      \mybottomrule
      \end{tabularx}}
    {\caption{Impact of temporal augmentation. See Table 1 nomenclature.}
      \label{tab_temporal_augm}}
  \end{floatrow}
\end{table}%

\textit{Domain-specific augmentation:} Table \ref{tab_color_augm} compares models trained with and without channel dropout on the Color Doppler input. Table \ref{tab_temporal_augm} compares results with and without temporal augmentation. Overall, we saw that the augmentations resulted in improved test accuracy, which suggests the model's robustness to the poor Doppler signal quality and varying freehand scanning motions.

\textit{Model size and baseline comparisons:} Comparing baseline UNet models modified to match the number of parameters as our final proposed models (Baseline-L, 0.3M parameters), a three-fold increase in parameters gave no appreciable improvement in performance (Table \ref{tab_summary}). This suggests that the improved accuracy of the context-aware models was not due simply to larger network size.

\textit{Cross validation:} To assess generalization, leave-one-out cross validation was carried out on both datasets using the contextually-aware models (VesNetSCT+). In each split, sequences from one subject were held out for testing (Table \ref{tab_cross_validation}).

\textit{Inference speed:} 
VesNetSCT++ achieved inference speeds of 149.4$\pm$4.6 ms (6.7 Hz) on a mobile CPU processor (Intel Core i7 2.6 GHz) and 8.9$\pm$0.6 ms (112 Hz) on a mobile GPU (Nvidia RTX 2080). These speeds were significantly faster than those of the Baseline-L UNet model and show that the proposed methods are amenable to real-time processing on hardware used in clinical US machines.


\begin{table}[!t]
\floatsetup{floatrowsep=qquad, captionskip=4pt}
\addtolength{\parskip}{-1em}
\setlength{\tabcolsep}{3.9pt}
{\caption{Cross-validation results with context-aware models (VesNetSCT+). 
}
  \label{tab_cross_validation}}
\begin{center}
\begin{tabular*}{\textwidth}{ccccccccc} 
 \mytoprule
  \makecell{\textbf{Data splits:}} & 
  \makecell{\textbf{1}} & 
  \makecell{\textbf{2}} & 
  \makecell{\textbf{3}} & 
  \makecell{\textbf{4}} & 
  \makecell{\textbf{5}} & 
  \makecell{\textbf{6}} & 
  \makecell{\textbf{7}} & 
  \makecell{\textbf{Mean $\pm$ Stdev}} \\
 \midrule
 \midrule
 Femoral arteries & \textbf{0.925} & 0.853 & 0.828 & 0.940 & 0.935 & 0.909 & 0.802 & \textbf{0.885 $\pm$ 0.162}\\
 Tibial arteries  & \textbf{0.671} & 0.782 & 0.565 & 0.511 & 0.646 & -     & -     & \textbf{0.635 $\pm$ 0.215}\\
 \mybottomrule
\end{tabular*}
\end{center}
\vspace{-0.55cm}
\end{table}
\addtolength{\parskip}{-0em}

\subsubsection{Visualization of Results.}
\begin{figure}
\includegraphics[width=\textwidth]{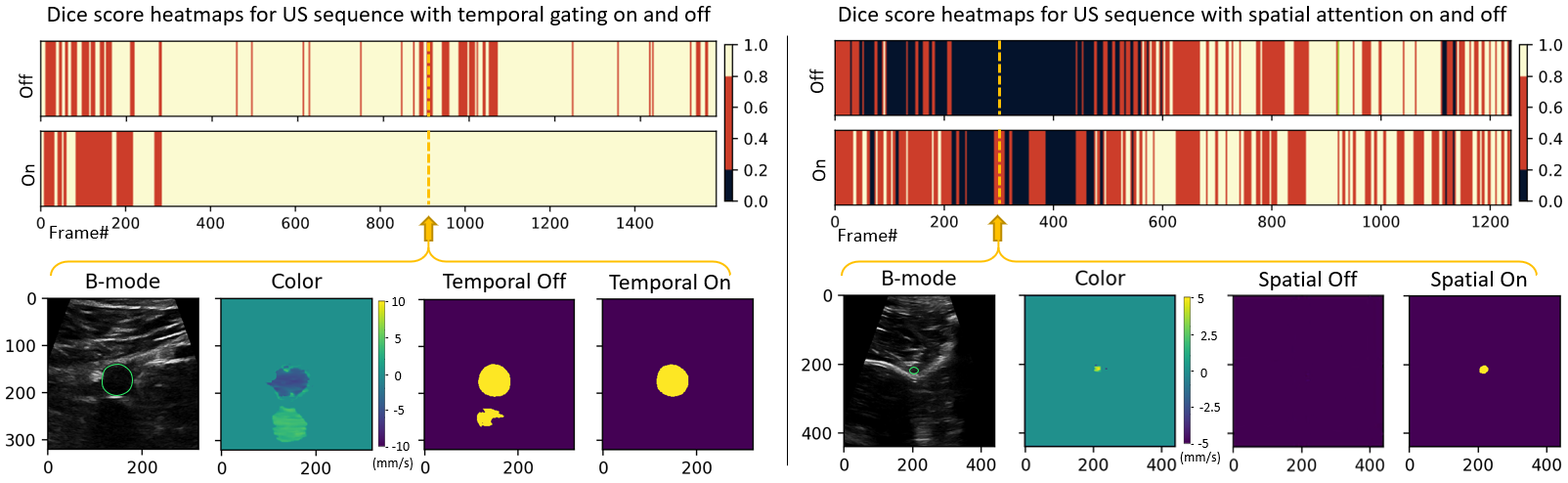}
\caption{Visualization of segmentations on hold-out test sequences. \textit{Left}: Temporally-aware model (VesNetT+) outperforms an equivalent model operating in individual frames (VesNet). \textit{Right}: Inclusion of spatial self-attention (VesNetST+) outperforms the same model without spatial attention (VesNetT+). 
} \label{fig_dice_curve}
\end{figure}

Fig. \ref{fig_dice_curve} shows examples of the effects of spatiotemporal context on hold-out sequences. In both cases, Dice scores fluctuate less when the contextual mechanisms are introduced. The addition of temporal gating (left panel, comparing VesNet and VesNetT+ models) allows the model to correctly learn pulsatile signatures and ignore confounding Doppler signals from nearby veins.
With spatial attention gating (right panel, comparing VesNetT+ and VesNetST+), extremely small tibial arteries are more reliably localized throughout.

\vspace{-0.25cm}

\vspace{-0.55cm}
\section{Conclusion}
This work presented an efficient deep learning architecture that incorporates multiple strategies for embedding spatiotemporal context to improve segmentation of challenging 2D US image sequences. 
We applied the methods to small-scale lower extremity arteries from freehand B-mode and Color Doppler scans, and showed strong improvement over baseline models without the added contextual awareness. Future work will investigate the generalizability of these methods on other anatomies where flow and spatiotemporal information are available, and where automatic quantification of vascular measurements is of clinical benefit.
\subsubsection{Acknowledgment.}  The authors thank Elizabeth Brunelle, Barbara Bannister, and Jochen Kruecker for assistance in data acquisition, annotation, and review.



%
%

\end{document}